# Demonstration of a monocrystalline GaAs-β-Ga$_2$O$_3$ p-n heterojunction

Jie Zhou[1,a)], Moheb Sheikhi[1,a)], Ashok Dheenan[2,a)], Haris Abbasi[1], Jiarui Gong[1], Yang Liu[1], Carolina Adamo[3], Patrick Marshall[3], Nathan Wriedt[2], Clincy Cheung[3], Shuoyang Qiu[1], Tien Khee Ng[4], Qiaoqiang Gan[4], Vincent Gambin[3], Boon S. Ooi[4,b)], Siddharth Rajan[2,b)], and Zhenqiang Ma[1,b)]

[1]*Department of Electrical and Computer Engineering, University of Wisconsin-Madison, Madison, Wisconsin, 53706, USA*

[2]*Department of Electrical and Computer Engineering, The Ohio State University, Columbus, OH 43210, USA*

[3]*Northrop Grumman Corporation, Redondo Beach, CA 90278, USA*

[4]*Department of Electrical and Computer Engineering, King Abdullah University of Science and Technology, Thuwal 23955-6900, Saudi Arabia*

a) These authors contributed equally to this work.
b) Author to whom correspondence should be addressed. Electronic mail: mazq@engr.wisc.edu, rajan.21@osu.edu, boon.ooi@kaust.edu.sa

In this work, we report the fabrication and characterizations of a monocrystalline GaAs/β-Ga$_2$O$_3$ p-n heterojunction by employing semiconductor grafting technology. The heterojunction was created by lifting off and transfer printing a p-type GaAs single crystal nanomembrane to an Al$_2$O$_3$ -coated n-type β-Ga$_2$O$_3$ epitaxial substrate. The resultant heterojunction diodes exhibit remarkable performance metrics, including an ideality factor of 1.23, a high rectification ratio of $8.04 \times 10^9$ at +/- 4V, and a turn on voltage at 2.35 V. Furthermore, at +5V, the diode displays a large current density of 2500 A/cm$^2$ along with a low ON resistance of 2 mΩ·cm$^2$.

**Key words**: semiconductor grafting, heterojunction, gallium oxide, gallium arsenide, transfer printing, nanomembrane

**Introduction**

Beta phase-gallium oxide (β-Ga$_2$O$_3$) is an emerging ultrawide bandgap material that shows significant interest in the field of power electronics and solar-blind optoelectronics.[1,2] This material has exceptional electronic properties such as a bandgap of around 4.9 eV, a high breakdown field of 8 MV/cm, and a high saturation velocity of $1.1 \times 10^7$ cm/s, rendering it ideal for high-power electronic applications.[3] In the meantime, the abundant raw material availability for β-Ga$_2$O$_3$, combined with the accessibility to large diameter bulk substrate, offers potential for cost-effective fabrication, positioning it as a competitor to other established wide-bandgap semiconductors like silicon carbide (SiC) and gallium nitride (GaN).

Despite its numerous advantages, further device development employing β-Ga$_2$O$_3$ still faces challenges. One of the primary concerns is its deficient p-type doping, which is crucial for creating bipolar devices.[4–6] Additionally, there exists a disparity in electron and hole mobilities, with reported values of 153 cm$^2$/V·s and 1.3 cm$^2$/V·s, respectively.[7,8] This pronounced difference in hole mobility can lead to asymmetrical charge transport properties, potentially constraining the performance of bipolar devices. In response to these challenges, hetero-integration strategies like wafer bonding[9–13], oxide deposition/sputtering[14–18], and van der Waals bonding[19,20], and heteroepitaxy[21,22] have been explored. While previous studies have significantly advanced the field of β-Ga$_2$O$_3$-based bipolar devices, the methodologies employed in these studies often encounter certain issues, including intermixing of atoms, polycrystalline nature of oxides, and possibly unpassivated dangling bonds at interfaces. Such challenges could impede the future progression of β-Ga$_2$O$_3$-based bipolar devices. Consequently, exploring innovative/alternative approaches that can potentially overcome some of, if not all, these obstacles would be needed in the pursuit of refining β-Ga$_2$O$_3$-based bipolar devices.

Semiconductor grafting emerges as a pioneering technology, enabling the formation of abrupt heterojunctions between dissimilar semiconductors, irrespective of their lattice constants or structural differences.[23–27] This method improves interfaces at semiconductor heterojunctions by incorporating an ultrathin oxide (UO) interlayer, typically sub-nanometer in thickness. This UO serves a dual purpose: it acts as a passivation layer, effectively suppressing the interfacial traps, and as a quantum tunneling layer, facilitating efficient charge carriers transport across the junction.

Leveraging this grafting technology, high-quality β-Ga$_2$O$_3$-based heterojunctions, such as Si/β-

$Ga_2O_3$ and $GaAsP/β-Ga_2O_3$, have been demonstrated.[28,29] Through the strategic introduction of either intrinsic oxide via chemical oxidation, or extrinsic oxide through atomic layer deposition (ALD) at interfaces, these pn junctions have displayed very low ideality factors, which are comparable to lattice-matched epitaxial-quality interfaces. In this study, we further extend this grafting strategy to showcase a high-performance heterojunction between GaAs and $β-Ga_2O_3$. This is achieved by employing a lift-off and transfer printing of a heavily doped p+ GaAs nanomembrane (NM) to an n-type $β-Ga_2O_3$ epitaxial substrate coated with an ultrathin layer of $Al_2O_3$, subsequently leading to the creation of heterojunction p-n diodes. The GaAs/ $β-Ga_2O_3$ pn diodes exhibit the following performance metrics: an ideality factor of 1.23, an impressive rectification ratio of $8.04 \times 10^9$, a turn-on voltage of 2.35 V, a large current density of 2500 $A/cm^2$, and a low ON resistance of 2 $mΩ·cm^2$. Beyond these remarkable device metrics, our work highlights several breakthroughs in $β-Ga_2O_3$ research. Primarily, we demonstrated the potential of semiconductor grafting in creating high-quality monocrystalline abrupt heterojunctions. Specifically, by utilizing grafting strategy, we technically addressed the intricate challenge of integrating GaAs with $β-Ga_2O_3$, a task complicated by significant discrepancy in lattice constants, thermal expansion coefficients, and thermal compatibility between the two semiconductors. Furthermore, with the successful integration of GaAs and $β-Ga_2O_3$, and by harnessing the high carrier mobility of GaAs and high breakdown field of $β-Ga_2O_3$ synergically, we anticipate the emergency of $β-Ga_2O_3$-based heterojunction devices of increased complexity and improved performance in more areas, for instance, optoelectronic devices such as solar-blind photodetectors and high-efficiency solar cells, and high-power and high-speed electronic devices such as heterojunction bipolar transistors (HBTs). We believe these insights could pave the way for unlocking the potential of this semiconductor material.

**Experiment and Results**

Fig. 1 (a) describes the synthesized heterostructure between GaAs and $β-Ga_2O_3$. The heterojunction is composed of a 400 nm thick, heavily doped GaAs layer, grafted to an n-type $β-Ga_2O_3$ substrate. The epi $β-Ga_2O_3$ layers contain a 250 nm thick lightly doped n- $β-Ga_2O_3$ layer, epitaxially grown on top of a 100 nm heavily doped n+ $β-Ga_2O_3$ layer. Both layers were epitaxially grown on an unintentionally doped (UID) intrinsic $β-Ga_2O_3$ bulk substrate. A comprehensive description of the epitaxy of $β-Ga_2O_3$ is available in the referenced literature.[29]

Sandwiched between the GaAs and β-Ga$_2$O$_3$ layers is an ultrathin oxide (UO) interlayer of Al$_2$O$_3$. This interlayer features the functionalities of double-side passivation and quantum tunneling, both of which are fundamental to semiconductor grafting.[23] Assuming a negligible potential drop across the Al$_2$O$_3$ interface, the band alignment of the GaAs/β-Ga$_2$O$_3$ heterostructure is depicted in Fig. 1 (b). Notably, the band offsets of conduction band and valence band are 0.07 eV and 3.41 eV respectively. The bandgap values for GaAs and β-Ga$_2$O$_3$ are 1.42 eV and 4.9 eV, respectively. Meanwhile, their electron affinities are 4.07 eV and 4.0 eV, respectively.

The fabrication of the GaAs/ β-Ga$_2$O$_3$ heterostructure started with the substrate preparation. First, the β-Ga$_2$O$_3$ epi (Fig. 2 (a)) was patterned using standard photolithography. Subsequently, it was dry etched using an inductively coupled plasma-reactive ion etching (ICP-RIE) process (Plasma Therm 770 ICP etcher, BCl$_3$: 18 sccm, Cl$_2$: 10 sccm, and Ar: 5 sccm, pressure 20 mTorr, RIE power 75 W, and ICP power 500 W), as depicted in Fig. 2 (b). This dry etching step was completed after exposing the underlying n+ β-Ga$_2$O$_3$ layer. The cathode trench regions were formed and were ready for the subsequent cathode metallization. In the cathode metal formation step, a metal stack of Ti/Au/Cu/Au (10/10/100/10 nm) was deposited on n+ β-Ga$_2$O$_3$ using an electron beam evaporator (Angstrom Engineering Nexdep Physical Vapor Deposition Platform). Afterwards, a metal lift-off process was performed, followed by an annealing process at 600 °C for 10 seconds to form ohmic contact, as presented in Fig. 2 (c). After completing the cathode metallization, the substrate underwent a standard cleaning procedure consisting of sonication in acetone, isopropyl alcohol (IPA), and deionized (DI) water for 10 minutes each, and a quick dip in buffer oxide etchant (BOE) for 10 seconds. Immediately after cleaning, the substrate was transferred into a custom ALD system, wherein 5 cycles of Al$_2$O$_3$ were deposited at 200 °C, as seen in Fig. 2 (d).

In parallel, the GaAs NM was obtained from a GaAs source epitaxial wafer consisting of a 400 nm $4 \times 10^{19}$ cm$^{-3}$ p+ GaAs layer sitting atop of a 300 nm underlying sacrificial AlAs layer, both epitaxially grown on a semi-insulating (SI) GaAs substrate. The fabrication of GaAs NM started with the patterning $9 \times 9$ µm$^2$ mesh holes and open areas for the cathode metal regions on p+ GaAs, followed by a dry etching process using the same ICP etcher (BCl$_3$: 10 sccm, and Ar: 5 sccm, pressure 15 mTorr, RIE power 60 W, and ICP power 500 W). This process exposed the underlying AlAs sacrificial layer. The dry etched GaAs epi was then soaked in diluted hydrofluoric acid (HF:H$_2$O = 1:20) for 20 minutes, which completely removed the AlAs layer, resulting in a freestanding p+ GaAs NM resting on the SI GaAs substrate. The p+ GaAs NM was then lifted using a polydimethylsiloxane (PDMS) stamp, followed by a rinse in a Tetramethylammonium hydroxide (TMAH) based developer (Microchem MIF-321) to remove the AlF$_3$ residues on the back side of the GaAs NM.[30]

The GaAs NM attached on the PDMS stamp was then precisely aligned and transferred to the cathode region-exposed β-Ga$_2$O$_3$ epi substrate using an MJB-3 aligner, as is schematically illustrated in Fig. 2 (e). After removing the PDMS stamp, the GaAs NM was weakly attached to the Al$_2$O$_3$-coated β-Ga$_2$O$_3$ epi substrate via van der Waals force. After performing a thermal anneal at 350 °C for 5 minutes using a rapid thermal annealer (RTA), a GaAs/Al$_2$O$_3$/β-Ga$_2$O$_3$ heterostructure with chemically bonded interface was formed, as shown in Fig. 2 (f).

The anode metallization was then performed on the p-type GaAs using a metal stack of Ti/Pt/Au/Cu/Au (15/50/10/100/50 nm) using standard photolithography and e-beam evaporation, as is shown in Fig. 2 (g). Then, a device isolation dry etching process was carried out by selectively etching away the exposed (not covered by the anode metal) GaAs NM regions, leading to a completed GaAs/β-Ga$_2$O$_3$ diode structure, as shown in Fig. 2 (h). After finishing the isolation step, all diodes were passivated with 80 cycles of Al$_2$O$_3$ (~8 nm) using the same ALD tool.

Figs. 3 (a) and (b) present atomic force microscopy (AFM) images of the as-grown n-/n+ β-Ga$_2$O$_3$ substrate and the GaAs NM post-transfer to β-Ga$_2$O$_3$, both characterized over a scanning area of 5 × 5 µm$^2$. The β-Ga$_2$O$_3$ substrate, post-epitaxial growth, exhibits an ultra-smooth surface with a root mean square (RMS) roughness of 0.714 nm. Meanwhile, the GaAs NM, even after going through the release and transfer process, maintained its high smoothness, with an RMS value of 0.873 nm. Fig. 3 (c) displays a differential interference contrast (DIC) image of the GaAs NM after being transfer printed to the β-Ga$_2$O$_3$ substrate. As can be seen, despite the large difference of the thermal expansion coefficient difference between GaAs NM and the β-Ga$_2$O$_3$ substrate, the GaAs NM, after undergoing thermal annealing, has been seamlessly bonded to the substrate, exhibiting excellent conformality and retaining its membrane integrity. Fig. 3 (d) provides a visual representation of a single GaAs/β-Ga$_2$O$_3$ p-n diode post-fabrication and electrical characterization. The left panel, Fig. 3 (d i), displays the device's DIC optical image, while the right panel, Fig. 3 (d ii), presents the scanning electron microscopy (SEM) image of a similarly sized device. Notably, anode, isolated GaAs NM, cathode, and exposed n+ β-Ga$_2$O$_3$ region are each indicated in both figures. Fig. 3 (e) illustrates the three-dimensional scanning image of a GaAs/β-Ga$_2$O$_3$ device, collected from an optical profiler using a Filmetrics®. This image illustrates a double-layered structure, highlighting an anode atop the GaAs NM and a cathode sitting on the n+ β-Ga$_2$O$_3$ layer.

The I-V characteristics of the synthesized diodes were evaluated using a Keithley 4200 Semiconductor Parameter Analyzer, with the results summarized in Fig. 4. The I-V profiles are delineated on linear scale in Fig. 4 (a), and semi-logarithmic scale in Fig. 4 (b). Both figures capture the I-V curves in various sweeping directions: the positive sweep (from -4 V to +5 V) is represented in red, while the negative sweep (from +5 V to -4 V) is marked in blue. Noteworthy, distinct behaviors have been observed under different sweeping directions. A notable shift in current zero-crossing points emerges: -1.4 V for positive sweep, and 0.2 V for the negative sweep. Such patterns have been reported in multiple β-Ga$_2$O$_3$-based Schottky barrier diodes (SBD) and p-n diodes.[16,31,32] This phenomenon can be attributed to the charging and discharging dynamics stemming from parallel capacitance during the measurement phase.

The key metrics are extracted from the negative sweeping curve. The ideality factor and rectification are determined to be 1.23 and $8.04 \times 10^9$ at +/- 4 V, respectively. Additionally, the diode's turn-on voltage is obtained by fitting the linear regime between 3V and 4.5V and extrapolating to a current of zero, leading to a value of 2.35 V, as illustrated in the I-V curve in Fig. 4 (a). At +5V, the current density (J) and ON resistance are recorded as 2500 A/cm$^2$ and 2 mΩ·cm$^2$, respectively.

In Fig. 5, we present a summary of representative Ga$_2$O$_3$-based diodes, including unipolar Schottky barrier diodes (SBD)[17,33] and bipolar heterojunction p-n diodes[12,16–18,20,22,28], benchmarked using metrics of ideality factor and rectification ratio. As can be seen from the figure, most heterojunctions formed between Ga$_2$O$_3$ and a foreign semiconductor, constructed through direct bonding or deposition, exhibit relatively high ideality factors and elevated leakage currents. In contrast, most oxide/Ga$_2$O$_3$ heterojunctions generally outperform, showcasing ideality factors below 2 and rectification exceeding seven orders of magnitude. Metal/Ga$_2$O$_3$ SBDs, being unipolar devices, consistently present near-unity ideality factors with high rectifications. Notably, our grafted semiconductor/Ga$_2$O$_3$ heterojunctions, specifically, Si/Ga$_2$O$_3$, and GaAs/Ga$_2$O$_3$, both show close-to-unity ideality factors of 1.13 and 1.23, respectively. Among these reported works, the grafted GaAs/Ga$_2$O$_3$ heterojunction stands out, registering a record-low diode ideality factor and a record-high setting of rectification ratio. This exceptional performance is attributed to the strategic incorporation of ultrathin oxide at the interface, which provides effective passivation on both surfaces of GaAs and $Ga_2O_3$ and facilitating quantum tunneling.

**Conclusion**

In conclusion, our work underscores the potential of semiconductor grafting technology in creating high-quality, lattice-mismatched monocrystalline β-$Ga_2O_3$-based heterojunctions. This approach offers a promising solution to the longstanding challenge of constructing β-$Ga_2O_3$ bipolar devices, as bottlenecked by the absence of efficient p-type doping. Furthermore, with the successful integration of GaAs and β-$Ga_2O_3$, unique strengths of each semiconductor can be harnessed. This synergy paves the way for constructing advanced β-$Ga_2O_3$-based electronic devices with boosted speed and power, as well as optoelectronic devices with enhanced efficiency and expanded functionalities.

**Acknowledgements**

The work was supported by a CRG grant (2022-CRG11-5079.2) by the King Abdullah University of Science and Technology (KAUST). The work also received partial support from DARPA H2 program under grant: HR0011-21-9-0109.

8

**Figures**

(a) 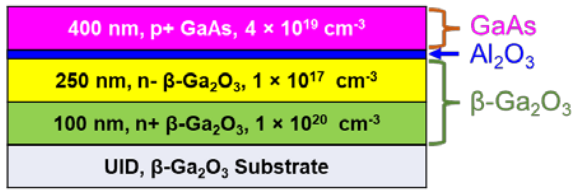   (b) 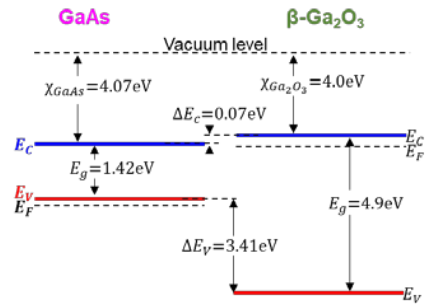

**Fig. 1.** (a) Schematic illustration of the GaAs/β-Ga$_2$O$_3$ p-n heterostructure. The heterojunction is formed between GaAs and β-Ga$_2$O$_3$ and interfaced by an ultrathin layer of Al$_2$O$_3$. (b) Band alignment of the GaAs/β-Ga$_2$O$_3$ heterostructure.

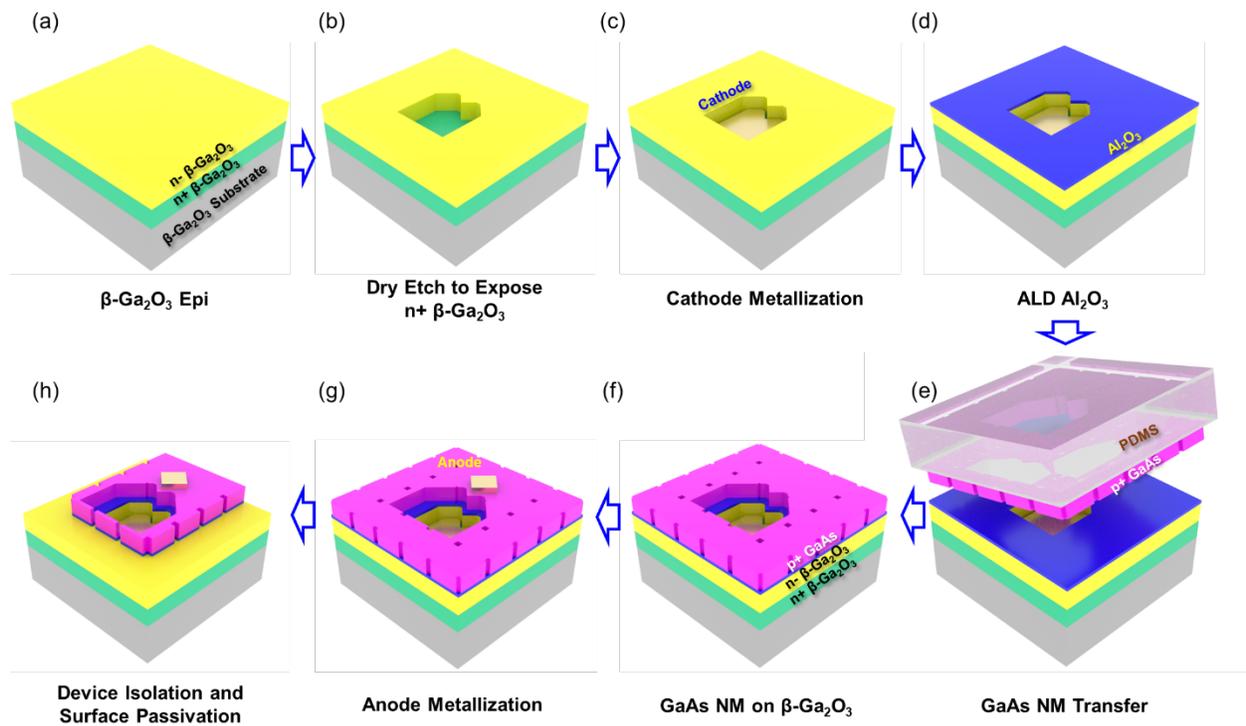

**Fig. 2.** Process flow illustration of GaAs/β-Ga$_2$O$_3$ heterostructure and p-n diodes. The entire process includes preparation of the β-Ga$_2$O$_3$ substrate, lift-off and transfer printing of the GaAs nanomembrane, grafting, and formation of the diodes.

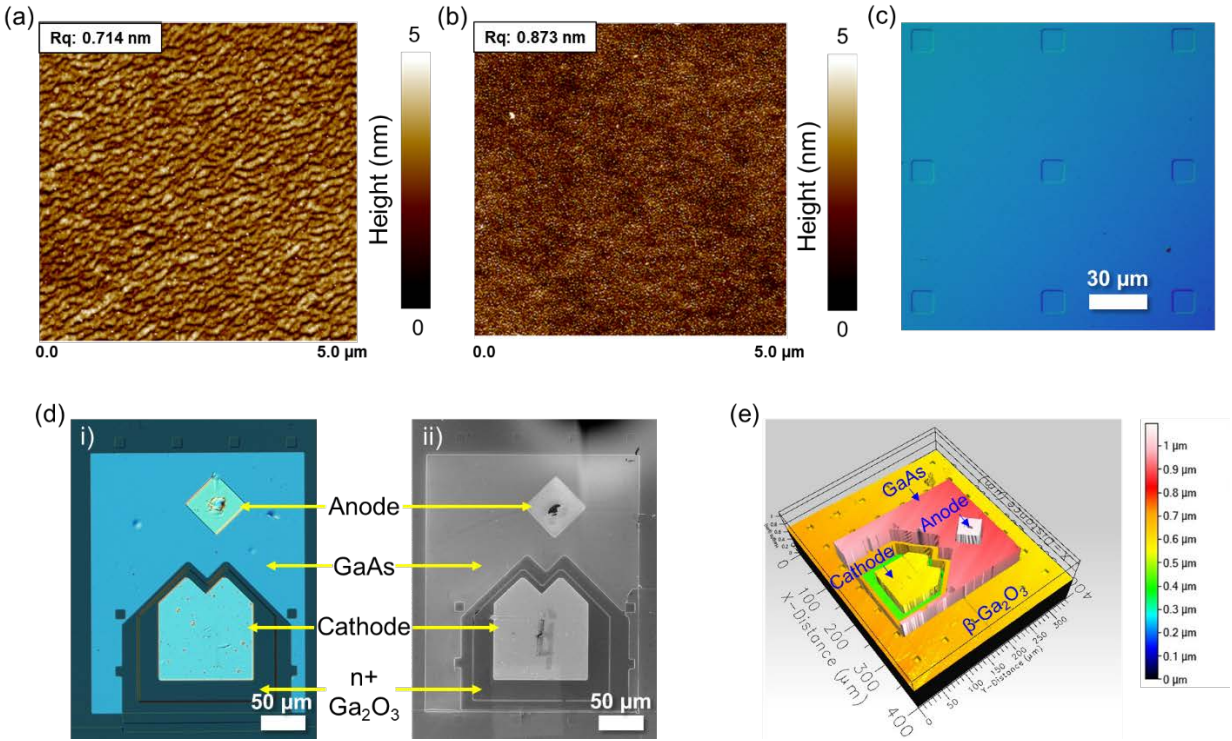

**Fig. 3.** Atomic force microscope (AFM) images of (a) as-grown β-Ga$_2$O$_3$ substrate, and (b) transfer printed GaAs nanomembrane. (c) Differential interference contrast (DIC) optical image of the GaAs nanomembrane transferred to β-Ga$_2$O$_3$ substrate. (d) Microscopic images of a singular fabricated device, with i) the optical DIC image, and ii) the scanning electron microscope (SEM) image. (e) A three-dimensional scanning image of the device captured through an optical profiler.

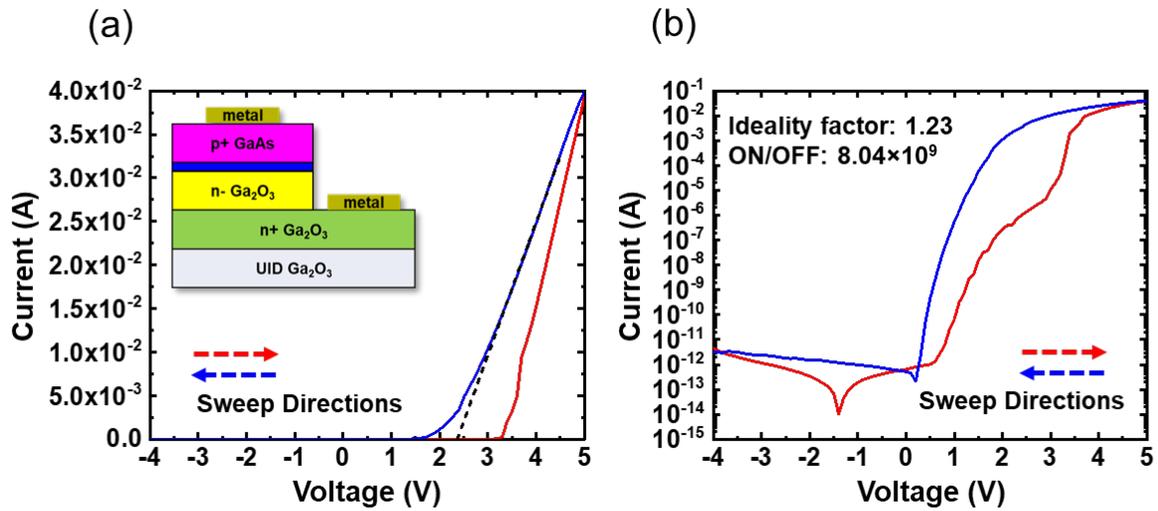

**Fig. 4.** Measured current-voltage characteristics of the GaAs/β-Ga$_2$O$_3$ heterojunction p-n diode, shown on (a) the linear scale, and (b) the semi-logarithmic scale. The red curves indicate the positive sweeping direction, while the blue curves represent the negative sweeping direction.

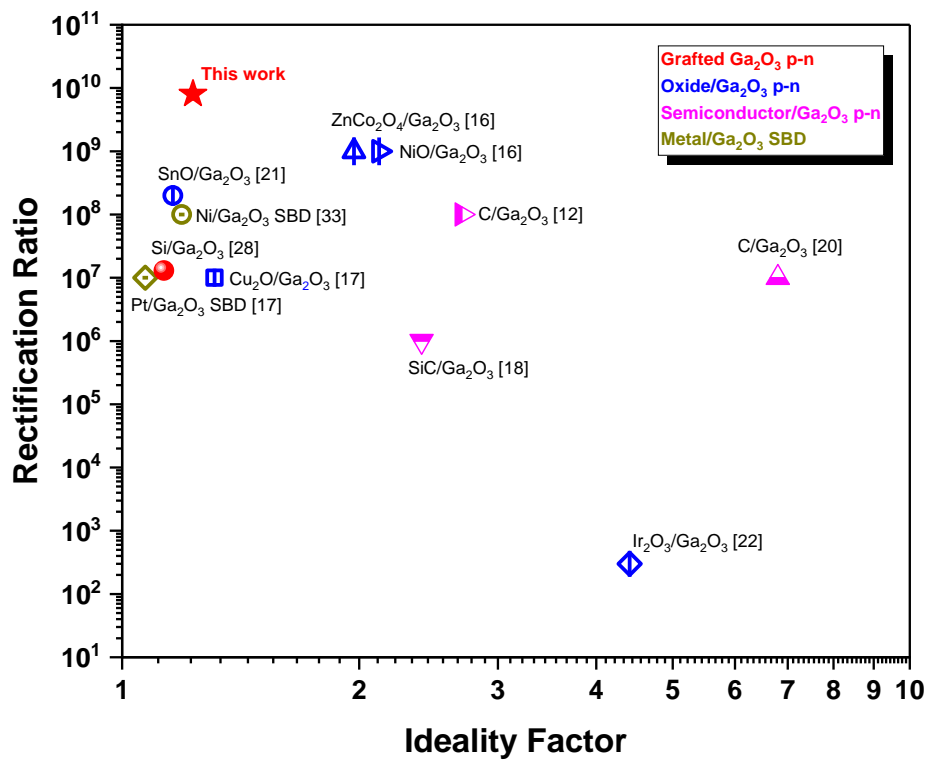

**Fig. 5.** Benchmark of the representative $Ga_2O_3$-based unipolar Schottky barrier diodes and bipolar heterojunction p-n diodes.